\documentclass[apjl]{emulateapj}

\newcommand\gsim{\,\lower3pt\hbox{$\sim$}\llap{\raise2pt\hbox{$>$}}\,}
\newcommand\lsim{\,\lower3pt\hbox{$\sim$}\llap{\raise2pt\hbox{$<$}}\,}

\shortauthors{LUGAZ ET AL.}
\shorttitle{BRIGHTNESS OF 3-D CMES}

\newcommand{\sSun}{ {\scriptscriptstyle{\rm \odot}} }

\begin{document}


%
%

\title{THE BRIGHTNESS OF DENSITY STRUCTURES AT LARGE SOLAR ELONGATION ANGLES:\\
WHAT IS BEING OBSERVED BY STEREO/SECCHI? }
\author{N.\ Lugaz \altaffilmark{1}, A.\ Vourlidas \altaffilmark{2}, I.~I.\ Roussev \altaffilmark{1}, C.\ Jacobs \altaffilmark{3}, W.~B.\ Manchester, IV \altaffilmark{4}, O.\ Cohen \altaffilmark{4}}
\altaffiltext{1}{Institute for Astronomy, University of Hawaii, 2680 Woodlawn Dr., Honolulu, HI 96822; nlugaz@ifa.hawaii.edu, iroussev@ifa.hawaii.edu}
\altaffiltext{2}{Code 7663, Naval Research Laboratory, Washigton, DC 20375; vourlidas@nrl.navy.mil}
\altaffiltext{3}{Centrum voor Plasma Astrofysica, KU Leuven, Celestijnenlaan 200B bus 2400, 3001 Leuven, Belgium ; carla.jacobs@wis.kuleuven.be}
\altaffiltext{4}{Department of AOSS, University of Michigan, 2455 Hayward St., Ann Arbor, MI 48198; chipm@umich.edu;oferc@umich.edu}
\submitted{Received 2008 May 23; accepted 2008 July 31}
\journalinfo{Astrophysical Journal Letters, unedited preprint}

%
%

\begin{abstract}

We discuss features of coronal mass ejections (CMEs) that are specific to heliospheric observations at large elongation angles.
Our analysis is focused on a series of two eruptions that occurred on 2007 January 24-25, which were tracked 
by the Heliospheric Imagers (HIs) onboard STEREO.  Using a three-dimensional (3-D) magneto-hydrodynamic simulation of these
ejections with the Space Weather Modeling Framework (SWMF), we illustrate how the combination of the 3-D nature of CMEs, solar
rotation, and geometry associated with the Thomson sphere results in complex effects in the brightness observed by the HIs. 
Our results demonstrate that these effects make any in-depth analysis of CME observations without 3-D simulations challenging. 
In particular, the association of bright features seen by the HIs with fronts of CME-driven shocks is far from trivial.  In this Letter, we
argue that, on 2007 January 26, the HIs observed not only two CMEs, but also a dense corotating stream compressed by the CME-driven
shocks.
 
\end{abstract}

\section{Introduction} \label{intro}

Coronal Mass Ejections (CMEs), one of the most extreme events in the solar system, have been studied extensively for over three decades now
\citep[see reviews by][]{Schwenn:2006,Roussev:2006}.  Until the launches of the {\it Solar Mass Ejection Imager (SMEI)} \citep[]{Eyles:2000} and the {\it Solar Terrestrial
Relations Observatory (STEREO)} \citep[]{Kaiser:2008} in 2003 and 2006, respectively, continuous observations of CMEs were made by the\textit{ SoHO\/}/LASCO coronagraphs
out to a radial distance from the Sun of 32~$R_\sSun$. 

The Heliospheric Imagers (HI-1 and HI-2, or collectively HIs)
\citep[]{Harrison:2005}, part of the {\it Sun-Earth Connection Coronal
  and Heliospheric Investigation (SECCHI)} suite onboard STEREO
\citep[]{Howard:2008}, enable the tracking of CMEs up to elongation
angles (i.e. angular distance from the Sun) of 90$^\circ$.   
These instruments observe the photospheric
light scattered by free electrons in the solar corona and heliosphere
\citep[Thomson scattering, e.g. see][]{Minnaert:1930}.  The
scattering strongly depends on the position of the observer relative to the radial direction of the observed electron. The scattering is maximized, along a given line of sight (LOS), at the point of closest approach to the Sun; namely, when the LOS is normal to the radial direction of the electron. The locus of these points for all possible LOS forms a sphere, called the Thomson sphere (TS) by \citet{Vourlidas:2006}, whose center is the halfway point between 
the Sun and the observer and whose diameter is equal to the Sun-observer distance. \citet{Jackson:1985} applied this concept to zodiacal photometric observations made by Helios at 
heliospheric distances as close as 0.31~AU. Recently,
\citet{Vourlidas:2006} examined the influence of the TS for the analyses
of CMEs observed by the HIs and SMEI by studying the evolution of the brightness of simple structures at large elongation angles. 

Additionally, the progenitors of Corotating Interaction Regions
(CIRs), which are dense streams associated with the interaction of a
fast stream of the solar wind originating from a 
coronal hole with the slow solar wind, have been imaged out to Earth's orbit
by the HIs \citep[]{Sheeley:2008b}.  \citet{Sheeley:2008a} and
\citet{Rouillard:2008} have described how the elongation versus time
tracks of the dense streams show apparent accelerations/decelerations
depending on the distance of the structures from the observer.  These
studies suggest the need for a revision of the standard paradigm of
tracking structures in the corona to adapt it to density structures
propagating in the inner heliosphere.  However, these revisions alone
are not sufficient to understand in detail the evolution of CMEs as
they propagate in the HIs' FOVs.  CMEs are intrinsically 3-D density
structures which, compared to CIRs, have a much larger and varying width
and can be rather asymmetric. The visibility of CME structures is, therefore,
subject to geometric and Thomson scattering effects that may be hard
to distinguish from evolution or propagation effects. To make
progress, fully 3-D numerical simulations of CME evolution in the
heliosphere associated with synthetic LOS capabilities are important \citep[]{Odstrcil:2005,
Lugaz:2005,Aschwanden:2006, Sun:2008}.

\citet{Manchester:2008} studied the evolution of CME brightness in the
HI-2's FOV and predicted two features due to the effects of the TS:
(i) the apparent deceleration (called ``stall'' by the authors) of a
CME front observed in HI-2 (similar to CIR observations by STEREO-A)
and, (ii) the appearance of a ``second'' front due to multiple
crossings of the TS by the CME.  However, their study was for an
extremely dense and fast CME (part of the Halloween storm); this
causes the solar wind background to be less important than for slower
CMEs.  In this Letter, we go further by analyzing a complex series of events, one
actually observed by STEREO: the 2007 January 24-25 CMEs interacting in
the heliosphere.  These were the first major CMEs detected by the HIs, and
they have been described by \citet{Harrison:2008}.  The
3-D modeling 
has been done using the Space Weather
Modeling Framework (SWMF) 
(\S \ref{Obs}). Here, we focus on the influence of the TS in
determining what density structures have been observed, focusing on
HI-2 observations on 2007 January 26 (\S \ref{LOS}).  
The final conclusions of this investigation are drawn
in \S \ref{conclusions}.

\section{THE SOLAR ERUPTIONS OF 2007 JANUARY 24-25} \label{Obs}

\begin{figure*}[ht*]
\begin{center}
{\includegraphics*[width=16cm]{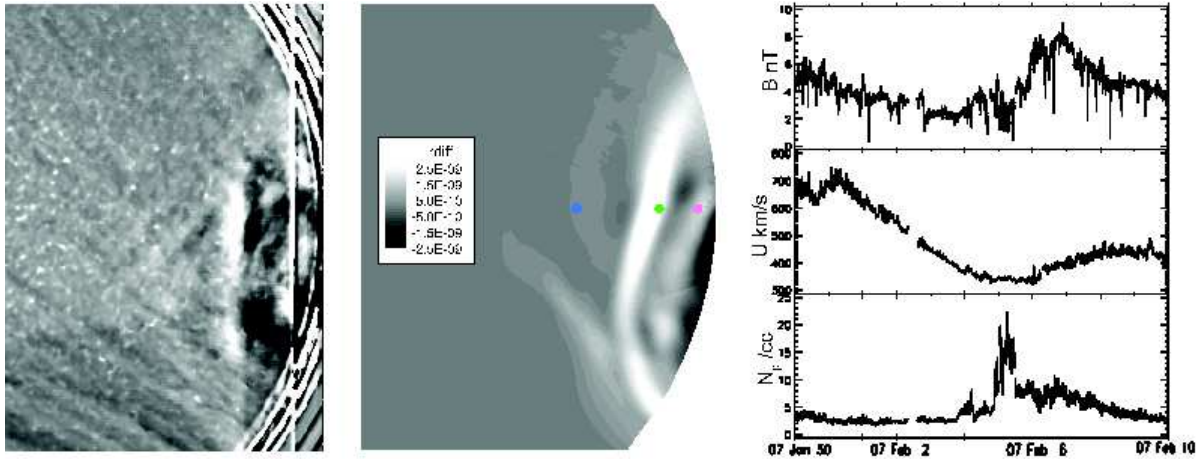}}
 \caption{{\it Left}: SECCHI HI-2A running difference image at 0600UT on 2007 January 26. {\it Middle}: Simulated SECCHI HI-2A running difference image at 0600UT on 2007 January 26.  The dots correspond to elongations of 30$^\circ$, 23$^\circ$ and 20$^\circ$ from left to right. {\it Right}: Wind observations during this time period showing a dense stream hitting the Earth on 2007 Feb 5, corresponding to the structure marked with the pink dot.}
\end{center}
\label{fig:6UT}
\end{figure*}

\subsection{Coronagraphic and Heliospheric Observations}\label{LASCO}
On 2007 January 24, at 1403UT, COR1 (FOV: 1.5-4~$R_\sSun$) observed an
eruption from active region (AR) 940, which was behind the eastern
limb of the Sun at the time.  At 0643UT on January 25, COR1 observed a
second eruption from the same AR. These two eruptions were also detected by SoHO/LASCO.
Based on time-height profiles in
the LASCO FOVs, they had a projected speed between 600 and
750~km~s$^{-1}$ for the first CME, and between 1,000 and
1,350~km~s$^{-1}$ for the second one, depending on the position angle
(PA, angle from the Solar North towards Solar East) of the measurement. Using these speeds derived from near-Sun observations and 
assuming that there was no deceleration, the two CMEs were
expected to interact on January 25 at 2030UT near PA 80$^\circ$ at a
radial distance of 97~$R_\sSun$ (0.45~AU).  Further assuming that both
CMEs occurred exactly on the east limb, the corresponding elongation
angle would be 25$^\circ$ (in HI-2's FOV).

At the time of the ejections, being in the commissioning phase of the mission, 
STEREO-A was rolled by 22.4$^\circ$ from
the solar north, which resulted in HI-1 and HI-2 imaging higher-latitude
regions than during the normal phase of the mission \citep[]{Harrison:2008}.  
The spacecraft was separated by approximately 0.4$^\circ$
from Earth and was at a radial distance of 0.97~AU from the Sun.  The
leading ejection was first detected in HI-1's FOV at 1801UT on January 24
at 4.4$^\circ$ elongation and was tracked until 12.1$^\circ$ at 0401UT
on January 25.  At that time, there was a SECCHI data gap for 20 hours.
Multiple fronts were detected in HI-2 on January 26: the brightest one
propagated from 24.7$^\circ$ at 0201UT to 32.5$^\circ$ at 1801UT.
\citet{Howard:2008} noted the presence of ``forerunner'' structures
(at 42$^\circ$ at 1801UT).  Even after accounting for the
deceleration of the CMEs, these observations are likely to have
covered the interaction of the two successive CMEs.  The ability of the HIs to
detect CME mergers in the heliosphere has been predicted by
\citet{Lugaz:2008}. The difficulty, however, is to associate the structures observed in HI-2 to the individual ejections.

\subsection{Simulation Set-up}\label{SWMF}
The 3-D simulation of the two ejections was done using the SWMF \citep[]{Toth:2005}.  
The solar corona domain of SWMF is
resolved with 40,489 blocks of $4^3$ cells each, ranging in size from
1/40~$R_\sSun$ at the inner boundary to 0.75~$R_\sSun$ at the outer
boundary.  The inner heliosphere domain of SWMF is resolved with
16,626 blocks of $8^3$ cells each, ranging in size from 3.44 to
0.215~$R_\sSun$. In both domains, the heliospheric current sheet is refined 
in order to better capture the density gradients there.
The solar wind and coronal magnetic field are reproduced using
the model developed by \citet{Cohen:2007}.  This model makes use of
solar magnetogram data and the Wang-Sheeley-Arge (WSA) model
\citep[]{Wang:1990, Arge:2000} for the asymptotic solar wind speed at 1~AU.  The
solar magnetic field is reconstructed from a Legendre polynomial
expansion of order 49 based on {\it NSO/SOLIS} magnetogram data.

To model the CMEs, we use a semi-circular flux rope prescribed
by a given total toroidal current, as in the models by
\citet{Titov:1999} and \citet{Roussev:2003b}.  A more complete
description of the implementation of this model can be found
in \citet{Lugaz:2007}.  This flux rope solution, once superimposed
on the background magnetic field, leads to immediate eruption
because of force imbalance 
The flux rope parameters are chosen such that there is an agreement with
the observed values of the speed of the eruptions in the corona. 
Our main goal here is to study the interaction of the two ejections with the
background solar wind and with each other.  

\section{Two CMEs and a Dense Stream} \label{LOS}

\subsection{Goal of the Study}
The STEREO spacecraft have great potential to enhance the 
quality of space weather predictions, because density structures can be tracked on their way to Earth
beyond LASCO FOV \citep[]{Kaiser:2008}. 
However, one of the difficulties associated with the interpretation of
SECCHI/HIs observations stems from the need to distinguish between
different types of density structures in the heliosphere.  
During the first year of the mission, observations focused on CIRs and
CMEs, which have been easily observed at large elongation angles.  
There has not been any reported observation of CMEs interacting
with CIRs in the HI-2's FOV yet.  This relative simplicity of HI-2's
observations should change as we approach the more active phase of
solar cycle 24.  Effects such as those discussed by
\citet{Vourlidas:2006}, \citet{Manchester:2008} and
\citet{Rouillard:2008} should make the direct association of bright
fronts observed by the HIs with ejections observed by coronagraphs
more complicated.  
In this section, we show how the knowledge gained from the careful
analysis of 3-D simulations can significantly improve the interpretation
of HI observations.  

On January 26, HI-2 onboard STEREO-A observed a number of structures
associated with the CMEs of 2007 January 24-25 (Fig.~1, left panel).
Here, we focus on the observations and model results at 0600UT on January 26 
to try to determine the origin of these structures. This particular time was chosen because three
distinct structures were clearly visible in HI2's image. In this analysis, we analyze running difference images, 
the most common type of imaging used to track faint structures far from the Sun. 

\subsection{The Two Ejections}

In the right panel of Fig.~2, we show a two-dimensional cut of the simulation results  as seen from the solar north pole (showing the density scaled by 1/R$^2$) 
at 0600UT on January 26 with the schematic of the observations. 
As can be seen in this image, the two CMEs have partially merged, 
but there are still two distinct fronts in the vicinity of the Sun-Earth line.
In our synthetic HI-2 image (Fig.~1, middle panel), there are two bright structures (blue and green
dots) associated with these two distinct ejections. The second front (green dot)
is brighter because it corresponds to material swept up with the first
ejection which has been additionally compressed by the second
ejection. This enhanced density is often observed
in sheaths associated with interacting CMEs \citep[]{Farrugia:2006, Lugaz:2008}.

The protruding front in the southern half of the synthetic HI-2 image
(also present, dimly, in the real HI-2 image) is also associated with
the leading edge of the first CME (blue dot). 
As seen in halo CMEs near solar minimum, the leading edge propagates 
faster in high-latitude regions, where the
Alfv{\'e}nic speed is higher and the density is lower than near the current sheet. 
Because STEREO-A was rolled by	 22.4$^\circ$ from the solar north, 
the center of the LOS images 
corresponds to mid-latitude regions. 
This roll enhances the apparent protrusion
of the first front. 

Following this analysis, we associate the brightest structure in the real HI-2
image 
with the CME on 2007 January 25, which is in the process of overtaking the previous CME. We believe 
that the fainter structure ahead corresponds to the first CME. We
will now focus on the origin of the third structure marked with a pink dot in the middle panel Fig.~1. 

\subsection{The Dense Stream (DS)}

In the left panel of Fig.~2, we show the running difference of the
3-D density on the Thomson sphere and the plane corresponding to the center of the LOS image (PA 67.6$^\circ$). This is an approximate way
to visualize the results of the simulation in a manner similar to the HI2 images. 
The three structures are marked with three LOS rays which correspond to the dots of the same color in the middle panel of Fig.~1.
Note that the Thomson scattering decreases quite slowly away from the TS \citep[]{Vourlidas:2006}. Therefore, density enhancements away from 
the TS should be considered, but are hard to visualize (see online animation). 
Looking at the simulation result, we have found that the pink dot is associated with a density increase
on the TS distinct from the two CME fronts, but the ray also
intersects the second CME.
After careful analysis of the simulation results, we have determined that this distinct density increase is likely associated with
a dense corotating stream (DS), which has been compressed by the two CMEs
(see Fig.~2, right). 

According to our simulation, the DS was present in the heliosphere prior to the CME passage (``DS'' in the right panel of Fig.~2), with a
density about 50$\%$ larger than the average density at the same
heliospheric distance.  The simulation shows how the DS is
compressed by the two CMEs (right panel of Fig.~2), and due to co-rotation it enters the
FOV of HI-2 in January 26.  
The presence of this density enhancement prior to the passage of the CMEs and the small angular span of this denser region
clearly rules out the possibility that this increased density is associated with the CMEs. 
Our results indicate that, in the absence of
any CMEs, the Earth would have passed a region of enhanced density
between 0900UT on Feb 4 and 0100UT on Feb 6 (peak density at 1700~UT
on Feb 4).  In reality, a DS was observed at the Earth on
Feb 5, as seen in the Wind data shown on the right panel of Fig.~1.
We should note that our simulation resolution at the Earth's location
is 3.44~$R_\sSun$, which corresponds to about 40
minutes for the solar wind data. Also, the timing given here is based
on the steady-state solution using synoptic (27-day) magnetograms.  We
believe these two effects explain in part the earlier arrival of the
DS at Earth as predicted from our solar wind model.
This leads us to believe that the DS is real and is associated with the density enhancement on the TS
at the location of the third structure. However, as noted above, the pink ray intersects
not only this DS but also the second CME.

\begin{figure*}[ht*]
\begin{center}
\includegraphics*[width=16.5cm]{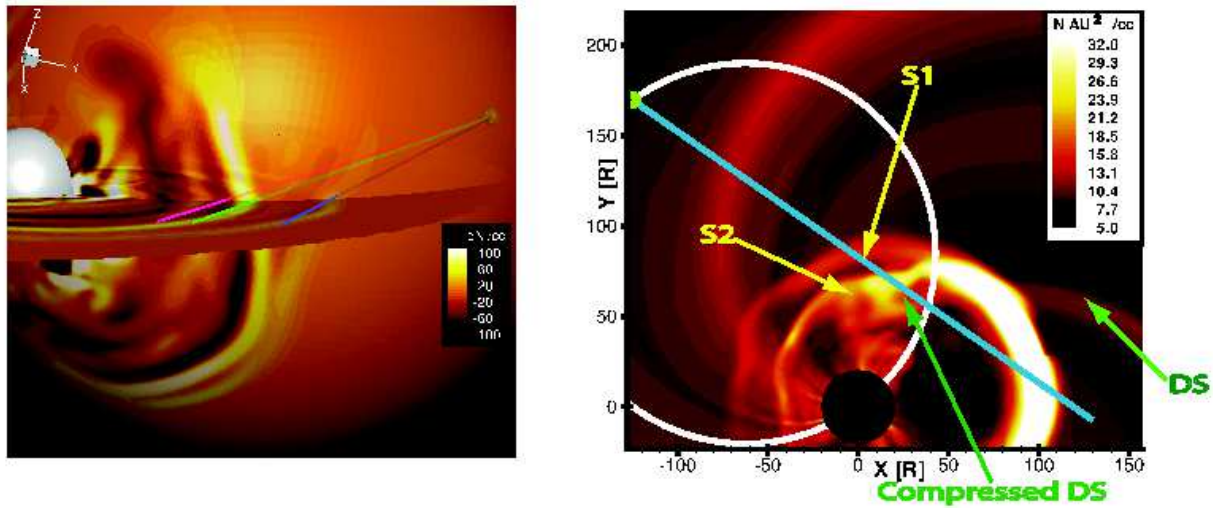}
\caption{{\it Left}: 3-D view of the CMEs at 0600UT on 2007 January 26 showing the density increase (dN) with respect to 0400UT.  The silver and green spheres depict the Sun and STEREO-A. 
The largest sphere is the TS. The three rays illustrate the LOS integration and correspond to the dots of the same color in the previous Figure. In the online version, the khaki and brown isosurfaces are regions
of dN = $\pm 80~cm^{-3}$ respectively.
{\it Right}: 2-D cut at 0600UT on January 26 as seen from solar north pole
   showing the density scaled by 1/R$^2$. The structures corresponding to the two CMEs are indicated with the S1 and S2 arrows. 
  The co-rotating dense stream before (resp. after) being hit by the CME fronts is indicated with the DS (resp. ``compressed DS'') arrow.
  The Sun, STEREO-A and the TS are represented by the black disk, the green disk and the white circle respectively. The blue line 
  delimits the inner FOV of HI-2A, 20$^\circ$ away from the Sun.}
\end{center}
\label{fig:DenseStream}
\end{figure*}


The analyzed HI-2 LOS image is a running
difference; therefore, a dense region along the pink ray will
correspond to a bright front at the pink dot only if no dense region
was present there in the previous image. In fact, the pink ray also
intercepted the CME front in the previous frame, whereas it did not
intersect the DS.  Hence, the CME front is not expected to contribute
to the brightness enhancement in the HI-2 image. Accordingly, we
believe that the brightness enhancement, in the running difference
image, is due to the DS, which was compressed by the two CMEs.

To summarize, there are three structures seen in the synthetic
LOS image: one corresponding to the leading CME having not
yet been overtaken by the following CME on the TS (blue dot); a second
one corresponding to the following CME, or the merged CMEs away from
the TS (green dot); and a third one resulting from the compression of
the co-rotating DS by the CMEs (pink dot).

Careful analyses of HI-2 images with the support of 3-D simulations such as the one presented here should help
develop a better understanding of which density structures at which position are tracked by STEREO HIs. This, in turn, will
help increase the accuracy of space weather predictions based on STEREO observations.

\section{DISCUSSION AND CONCLUSIONS}\label{conclusions}
In this Letter, we have discussed some apparent difficulties associated
with the interpretation of STEREO HI-2 images, especially when there
are multiple density structures in the field-of-view of the
instrument. By means of a fully 3-D MHD simulation of the 2007 January 24-25
ejections, we have been able to associate the observed HI-2 features with
features seen in the synthetic HI-2 images. We have found that they
correspond to the front associated with the leading CME, the front
associated with the newly merged CMEs, and a co-rotating DS
compressed by the two ejections. We believe that the observations could have been 
misinterpreted without the aid of a numerical simulation. This is because the DS appearance and evolution 
could have been mistaken for those of a CME. This work
emphasizes the importance of carrying out 3-D simulations of CME
interaction with a structured solar wind in interpreting STEREO
observations.
It should also be remembered that, without carefully considering the effect of subtracting consecutive images, 
one might not be certain with which structure (and which part of a structure) a bright front is associated. This is because
most observational rays intersect multiple density structures in the heliosphere.  
In future related studies, we will present a detailed comparison of the simulated
and real SECCHI images for the 2007/01/24-25 events, thus filling the 20-hour SECCHI data gap. 

\acknowledgments
  The simulation reported here was carried out on a dedicated cluster of the solar group at
  the IfA. The research for this
  manuscript was supported by NSF grants ATM0639335 and ATM0819653 and NASA grants
  NNX07AC13G and NNX08AQ16G. W.~M. was
  supported by NASA grant NNX06AC36G. We would like to thank the reviewer for helping us improve and clarify this Letter.
  The SECCHI data is produced by an
  international consortium of the NRL, LMSAL and NASA GSFC (USA), RAL 
  and U. Bham (UK), MPS (Germany), CSL (Belgium), IOTA and IAS (France).
  SOLIS data were obtained from NSO.

\end{document}